\documentclass[conference]{IEEEtran}
\IEEEoverridecommandlockouts
 
\usepackage{cite}

\usepackage{balance}

  \usepackage[pdftex]{graphics}
\ifCLASSINFOpdf

\else
\fi

\usepackage{amssymb} 
\usepackage{graphicx}
\usepackage[usenames,dvipsnames]{color}

\usepackage{url}

\usepackage{amsmath}
\usepackage{textcomp} 
\usepackage{xspace} 
\usepackage{bbding} 
\usepackage{pifont} 
\usepackage{mdwlist} 

\newcommand{\Prop}{\mathbb{PROP}}
\newcommand{\PropL}[1]{\mathbb{LPROP}^{#1}}
\newcommand{\PropA}[1]{\mathbb{ABSTR}^{#1}}
\newcommand{\PropAK}[1]{\mathbb{ABSTR_{KNOW}}^{#1}}
\newcommand{\Tests}[1]{\mathbb{TESTS}^{#1}}

\newcommand{\EnvAsm}[1]{\mathbb{ENV_{ASM}}^{#1}}

\newcommand{\E}{{\cal E}}

\begin{document} 

\title{Towards a Human-Centred Approach in \\ Modelling and Testing of Cyber-Physical Systems}

\author{
\IEEEauthorblockN{
Maria Spichkova\IEEEauthorrefmark{1},
Anna Zamansky\IEEEauthorrefmark{2}, 
Eitan Farchi\IEEEauthorrefmark{3}
}

\IEEEauthorblockA{\IEEEauthorrefmark{1} RMIT University, Australia,  
maria.spichkova@rmit.edu.au
}
\IEEEauthorblockA{\IEEEauthorrefmark{2} University of Haifa, Israel, annazam@is.haifa.ac.il
}
\IEEEauthorblockA{\IEEEauthorrefmark{3}  
IBM Research Lab, Haifa, Israel, farchi@ibm.co.il
}

}

\maketitle

\begin{abstract} 
The ability to capture different levels of abstraction in a system model is especially important
 for remote integration, testing/verification, and manufacturing of cyber-physical systems (CPSs).  
 However, the complexity of modelling and testing of CPSs makes these processes extremely prone to human error.
  In this paper we present our ongoing work on introducing {\em human-centred} considerations into modelling and testing of CPSs, which allow for agile iterative refinement processes of different levels of abstraction when errors are discovered or missing information is completed.   
\end{abstract}


\IEEEpeerreviewmaketitle

 \section{Introduction}
 
An appropriate system model provides a better overview as well as the ability to fix more inconsistencies more effectively and earlier in system development life cycle, reducing overall effort and cost. 
Nevertheless, modelling assumes abstraction of several aspects, especially the modelling of \emph{cyber-physical systems} (CPSs) on the level when we represent physical components and the corresponding properties. 
Even a very precise model cannot fully substitute for a real system. 
Many approaches on CPSs 
omit an  abstract logical level of the system representation and lose the advantages of the abstract representation. 
Some researchers \cite{Sapienza2690,Spichkova_Campetelli2012} 
suggested using a platform-inde\-pendent architectural design in the early stages of system development. 
The approach presented by Sapienza et al. \cite{Sapienza2690} introduces the idea of pushing
hardware- and software-dependent design  as late as possible.  
In comparison to \cite{Sapienza2690}, 
the focus of \cite{Spichkova_Campetelli2012}  
is on adaptation and generalisation of the software development methodologies.   
In our work, we extend these ideas by combining integration of quality-oriented aspects into the architectural levels with integration of {\em human-oriented} aspects into the process of system testing. 

In the context of quality-oriented aspects, we propose a testing methodology 
for CPSs which integrates different abstraction levels of the system representation. 
The crucial points for each abstraction level are $(i)$  whether we really require the whole representation of a system to analyse its core properties, 
and $(ii)$ which test cases are required on this level. 
In many cases, it is enough to represent some parts of the system that are relevant to a concrete purpose. This approach is based on the idea of refinement-based development of complex, interactive systems \cite{spichkova2008refinement,ArchReqDecRef,broy_refinement,broy_janus,spichkova2010architecture}. 

The above refinement can be thought of as {\em static}: once the abstraction levels have been determined, they remain fixed throughout the test planning process. In practice, however, due to the complex and error prone character of modelling and test planning, the modeller/tester often makes mistakes and may revisit the different levels of abstractions to make {\em dynamic} refinements. An underlying theory is therefore needed to understand how such dynamic refinements of one level of abstraction affects the other levels. This theory can be then used as a basis for developing tools supporting the human modeller/tester in such refinements. 

Our aim, therefore, is to extend our previous work on 
remote cyber-physical integration/interoperability testing \cite{Liu2015ASWEC1,Liu2015ASWEC2} 
by introducing human-centric elements into it, along the lines of \emph{Human-Centred Agile Test Design} (HCATD, cf. \cite{Zamansky2015HOFM}). This is particularly important as, by the \emph{Engineering Error Paradigm}~\cite{RedmillRajan},  
humans are seen as they are almost equivalent to software and hardware components in the sense of operation with data and other components, but at the same time humans are seen as the ``most unreliable component'' of the total system. 
Thus, in the case of testing of a CPS the  ``most unreliable component''  would be the tester. 
The Engineering Error Paradigm suggest designing humans out of the main system actions through automatisation 
of some system design steps is considered as a proposal for reducing risk. 

The main idea of HCATD is explicitly acknowledging that that the tester's activity is not error-proof: 
errors can happen, both in the model and the test plan, and should be taken into account. 
More concretely, discovery of an error or incomplete information may cause the tester to return to the model and refine it, which in its turn may induce further changes in the existing test plan. The term `agile' is meant to reflect the iterative and incremental nature of the process of modelling and test planning.

\emph{Contributions:} 
 We propose a human-centered agile modelling and testing approach for cyber-physical systems, which combines two types of refinements: static (or system-oriented, meant to hide unnecessary details) and dynamic (or tester-oriented, meant to provide the ability to correct and complete the developed artefacts). Developing appropriate tools for supporting this new paradigm may increase efficiency of testing of CPSs and reduce the testing cost and time by following the agile paradigm.


\section{Remote testing of CPSs}
\label{levAbs} 

A crucial question for a quality-oriented architecture in a global context is which features we need to check at which level of abstraction. 
Testing, as well as verification,  at  the concrete level is more expensive than on an abstract one, especially if some corresponding corrections within the system are necessary. 
Thus, it makes more sense to have more intensive testing at logical level to reduce the overall size of test suite for the next levels as much as possible.  
In \cite{spichkova2015SAGRA} we have suggested to have three main meta-levels of abstraction:  
\begin{itemize} 
\item 
\emph{Abstract  Level}, where we operate on the logical architecture of the system and an abstract model of the environment, and test. 
 the interoperability between logical components of our architecture; 
\item
\emph{Virtual Level}, where distinguish software and hardware architectures and
operate on both virtual and real representations of the hardware components.  
On this level we test the interoperability between virtual and real systems.
\item
\emph{Cyber-Physical Level}, where we operate on real system components, and test 
 the interoperability between real systems that are physically present for testing.
\end{itemize}
One of the advantages of this approach is the conformity with the ideas of \emph{Virtual Commissioning}  technology \cite{Drath2008,makris2012virtual}, 
which promises a more efficient handling of the complexity in assembly systems. 
Another advantage is the conformity with the top-down development methodology for the development of safety-critical software, especially for the automotive domain, cf. \cite{feilkas2011refined,feilkas2009top,VerisoftXT_FMDS}. 
In our current work we apply the HCATD methodology on each abstraction level, taking into account the error-prone nature of the human tester's tasks, 
and supporting the tester in refining and optimising test sets on each abstraction level. 
To increase the productivity on the Virtual Level, we can use facilities as the Virtual Interoperability Testing Laboratory~(VITELab, cf. \cite{blech2014cyber, issec2013spichkova}), where
the interoperability simulation and testing are performed early and remotely.  
At some level we need to switch from the pure abstract (logical) representation of the system to a cyber-physical one, but during a number of refinement steps we 
test (and refine) the system or component using a virtual environment, and then continue with testing in a real environment. 

When we mark some system properties as too concrete for the current specification layer and 
omit them to increase the readability and the understandably of the model, we have to check whether any important information about the system might be lost due this omittance, on this level of abstraction or in general. 
When we mark some system tests as unnecessary/optional to increase efficiency of the testing process, we have to check whether some important system properties are not covered by the chosen test set. 
Thus, on each abstraction level the traceability between the system properties and the corresponding tests is crucial for our approach.

If the information is not important on the current level, it could influence on the overall modelling result after some refinement steps, i.e., at more concrete levels that are more near to the real system in the physical world. Therefore, specifying system we should make all the decisions on abstraction in the model transparent and track them explicitly -- in the case of contradiction between the model and the real system this allows to find the problem easier and faster.

\begin{figure*}[ht]
\begin{center}
\includegraphics[scale=0.55]{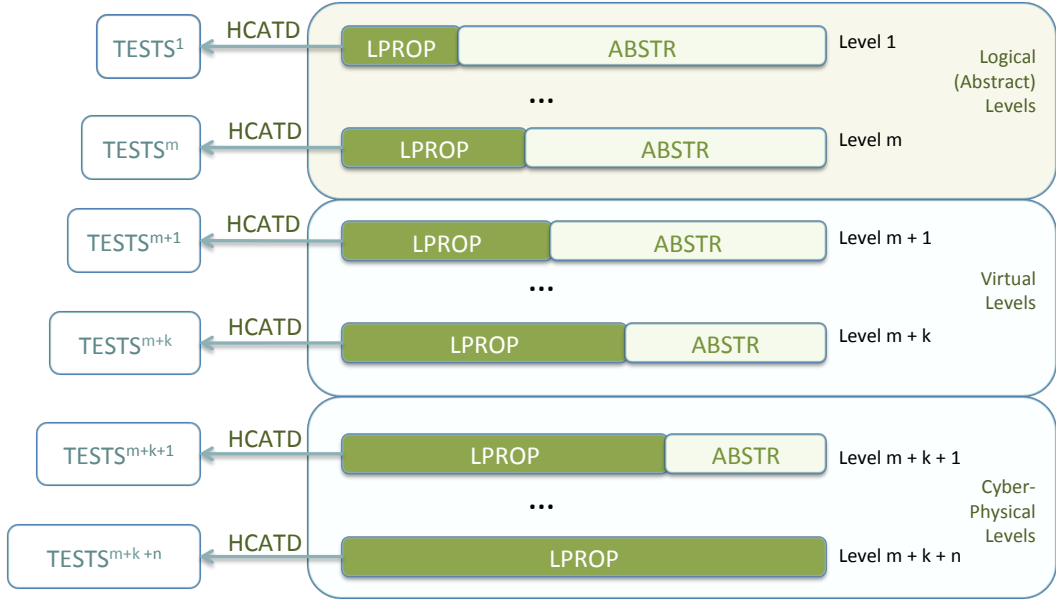}
\end{center}
\caption{Abstraction Levels of the Agile Combinatorial Test Design}
\label{fig:sbstrLevels}
\end{figure*}


\section{Formal CPS Testing Framework}
\label{framework} 

In general, we can say that any system $S$ can be completely described by the set $\Prop(S)$ of its (cyber-physical) properties. 
 On each level $l$ of abstraction we can split $\Prop(S)$ into two subsets: set  $\PropL{l}(S)$ of the properties reflected at this level of abstraction, and set $\PropA{l}(S)$ of properties from which we abstract at this level, knowingly or unknowingly. 
 We denote by $\PropAK{l}(S)$ the properties of the system from which we abstract intentionally and which we aim to track during system development, 
$\PropAK{l}(S) \subseteq \PropA{l}(S)$. 

 For any abstraction level  $l$ the following holds:\\
 $\PropL{l}(S) \cup \PropA{l}(S) = \Prop(S)$ and\\
  $\PropL{l}(S) \cap \PropA{l}(S) = \emptyset$. 
  
 Each property $p \in  \PropL{l}(S)$ should be covered by the corresponding tests on the level $l$. 
On the other hand, we do not need to specify on this level any tests to cover the properties from the set $\PropA{l}(S)$. 
Thus, we can say that the set $\Tests{l}(S)$ of tests required on the level $l$ have to be generated from the set $\PropL{l}(S)$ (cf. also Figure~\ref{fig:sbstrLevels}).  

With each refinement step we move some part of system's properties from the set $\mathbb{ABSTR}$ to the set $\mathbb{LPROP}$. 
We can say that in some sense the set $\mathbb{ABSTR}$  represent the termination function for the modelling process: 
in the case $l$ corresponds to the real representation of the system, we get $\PropL{l}(S) = \Prop(S)$ and $\PropA{l}(S) = \emptyset$. 

On each level $l$ we use a number of assumptions on environment of a system $S$.
We denote this set of assumptions by $\EnvAsm{l}$. 
In practice, we view the abstraction levels as corresponding to stages in an imperfect process rather than views which are kept complementary and consistent.
In comparison to the sets $\PropAK{l}$, it is unrealistic to expect monotonicity between the number $l$ and the cardinality of the set $\EnvAsm{l}$:
some assumptions on the environment could become weaker or unnecessary with the next refinement step,
but for some assumptions stronger versions may be needed or the system can require some new assumptions in order to fulfil all its properties.
However, it is important to trace the changes of $\EnvAsm{}$ on each level of modelling to find out which properties of the model on which levels should be re-tested,
if on some refinement step $l+1$ a contradiction between the $\EnvAsm{l}$ and the real targeting environment will be found.
Thus, the collected assumption should be checked during the testing phase, and if something is missed or incorrect, the model should be changed accordingly to the results of the testing.

\section{Introducing Human-Centred Aspects into CPS Testing Framework}

An agile software development process (ASDP) focuses on facilitating early and fast production of working code \cite{Turk2005Agile,Rumpe2006Agile,hazzan2014agile}. 
The corresponding ASDP models have support iterative, incremental development of software.
ASDP requires agile testing practices and guidelines, cf. e.g., \cite{hellmann2012agile,talby2006agile}.

The idea of an \emph{human-centered agile test design} (HCATD) was first introduced in the context of combinatorial test design   in \cite{Zamansky2015HOFM}.   
Combinatorial test design (CTD, cf. \cite{zhang2014introduction,segall2012common,farchi2014combinatorial,farchi2013using}) is an effective test planning technique, in which the space to be tested, called a {\em combinatorial model}, is represented by a set of parameters, their respective values and restrictions on the value combinations. The main challenge of CDT is to optimise the number of test cases, while ensuring the coverage of given conditions.
Tai and Lei \cite{Tai} shown in their experimental work that a test set covering all possible
pairs of parameter values can typically detect 50-75\% of the bugs in a program.
Kuhn et al.  \cite{Kuhn:2004} proved that typically all bugs can be revealed by covering the interaction
of 4-6 parameters. 
CTD approaches can be applied at different phases and scopes of testing, including end-to-end and system-level testing and feature-, service- and application program interface-level testing.  
In \cite{spichkova2015SAGRA} we have suggested to apply an architecture-based combinatorial testing approach for testing of CPSs, with the aim to increase the architectural sustainability, in the sense of cost-effective longevity and endurance, especially from the perspectives of integration in
a global context. In our current work we are going further bringing the human-centred agile aspects to the development process.

In CTD a system is modeled using a finite set of system parameters ${\bf P} = \{{\bf A}_1,\ldots,{\bf A}_n\}$ together with their corresponding associated values $\{{\bf V}(\bf A_1),\ldots,{\bf V}({\bf A_n})\}$. Our interest is in {\em interactions} between the different values of the parameters, i.e., elements of the form $\mathcal I\subseteq\bigcup_1^m {\bf V}({\bf A}_i)$, where at most one value of each parameter may appear. An interaction of size $n$ (where some value of each system parameter appears) is a {\em scenario} (or {\em test}). We say that a set of scenarios $T$ {\em covers} a set of interactions $C$ if for every $c\in C$ there is some $t\in T$, such that 
$c\subseteq t$. 

A {\em combinatorial model} $\E$ of the system is a set of scenarios, which defines all tests executable in the system. A {\em test plan} is a triple 
$P = (\E,C,T)$, where $\E$ is a combinatorial model, 
$C$ is a set of interactions called coverage requirements, and $T$ is a set of scenarios called tests, where $T$ covers $C$. 

One of the most standard coverage requirements is {\em pairwise testing} \cite{Nie:2011,Tai}: considering every (executable) pair of possible values of system parameters. In the above terms, a pairwise test plan can be formulated as any pair of the form $P = (\E,C_{pair}(\E),T)$, where $C_{pair}$ is the set of all interactions of size 2 which can be extended to scenarios from $\E$. 

Typically, the CTD methodology is applied in the following stages. First the tester constructs a combinatorial model of the system by providing a set scenarios which are executable in the system. 
After choosing the coverage requirements, the second stage is constructing a test plan, i.e., proposing a set of tests over the model, so that full coverage with respect to a chosen coverage strategy is achieved. 
More concretely, the goal of the tester is to provide a valid test plan $P = (\E,C,T)$. By the \emph{validity property} we mean here that 
\begin{itemize}
\item[(i)] $T$ satisfies all coverage requirements in $C$, and 
\item[(ii)] $T$ is a subset of $\E$. 
\end{itemize}
However, errors are possible at both of these stages, especially because of the human factor\cite{mioch2011selecting,spichkova2015human,Walia2013,Dhillon}. 
 Moreover, error discovery in the test plan may cause the tester to return to the model and refine it, and vice versa. The proposed term `agile planning' reflects the iterative and incremental nature of these two stages, based on the assumption that errors can happen, both in the model and the test plan. Therefore, neither {\em  correctness} nor {\em completeness} of the combinatorial model is not assumed at the stage of test planning, and the tester may go back to refining the model at any point.

While in standard CTD approaches it is assumed that (i) is satisfied before handling (ii), in HCATD they are handled {\em in parallel}. Thus in our framework we do not assume its availability (and correctness). Rather it is extracted iteratively when the tester specifies a set of specific test cases $T$, as well as some logical restrictions (in the form of formulas as defined above) on the combinatorial model, which provide only partial information about $\E$. For this reason uncertainty is explicitly represented in our framework by dividing the space of tests into three basic types, according to the information available from the tester:
\begin{itemize}
\item[(a)] 
\emph{Validated}:  the tester confirmed these tests as executable (according to some chosen confirmation strategy); 
\item[(b)]
 \emph{Rejected}: the tester rejected these tests as impossible or irrelevant on the current abstraction level; and 
\item[(c)] 
\emph{Uncertain}: the tester has not classified these tests to be validated/rejected, as not enough information has been
provided for the classification.
\end{itemize}
The final goal is a minimization of uncertainty by posing a series of queries to tester aiming to confirm/reject tests. 
\\

\noindent
\emph{Example Scenario:}  
Let us consider a system in which there
are two robots $R_1$ and $R_2$ interacting with each other. Suppose that on the abstraction level 0, the system is modeled as follows. 
Each robot has a gripper which has a mode ($GM_1$ and $GM_2$) -- either open or closed (to hold an object). 
Each robot is in one of the specified positions ($P_1$ and $P_2$), which is abstracted by the finite set of possible values $\{pos^1,pos^2,pos^3\}$ (on further meta-level more detailed positioning of the grippers might be provided). 

The most standard coverage requirement in the domain of combinatorial test design is pairwise testing\cite{Nie:2011,Tai}
  considering every (executable) pair of
possible values of system parameters.
Thus, let suppose the coverage requirement is pairwise coverage for the example scenario. 
We specify two meta-operation $Give$ and $Take$ to model the scenario when one robot hands an object to another robot. 
A meta-operation \emph{Give} in which $R_1$ gives an object to $R_2$ can only be performed when the grippers of both robots are in the same position, the gripper of $R_1$ is closed and the gripper of $R_2$ is open. 
This means that not not all test cases are executable and further restrictions should be imposed. 
Suppose, however, that the tester erroneously omits the information on the position, providing only the logical condition $GM_1= close$ and $GM_2 = open$ for the system model.

\begin{table}
 \caption{Example scenario: System model}
   \label{tab:example}
 \begin{center}
\begin{tabular}{|c|c|c|c|}
\hline 
$P_1$ & $P_2$ & $GM_1$ & $GM_2$ \\ \hline
\hline 
pos$^1$ & pos$^1$ & close & open  \\ 
\hline 
pos$^1$ & pos$^2$ & close & open \\ 
\hline 
pos$^1$ & pos$^3$ & close & open \\ 
\hline 
pos$^2$ & pos$^1$ & close & open  \\ 
\hline 
pos$^2$ & pos$^2$ & close & open \\ 
\hline
pos$^2$ & pos$^3$ & close & open \\ 
\hline
pos$^3$ & pos$^1$ & close & open  \\ 
\hline 
pos$^3$ & pos$^2$ & close & open \\ 
\hline  
pos$^3$ & pos$^3$ & close & open  \\ 
\hline
pos$^2$ & pos$^2$ & close & open \\ 
\hline

\end{tabular}
\end{center}
\end{table}

This induces a system model (cf. Table \ref{tab:example}), in which all tests may be marked as \emph{uncertain}, i.e., not yet confirmed by tester. 
The tester then goes on to construct a set of tests $\{t_1,t_2\}$, where\\ 
$t_1 = \{P_1:pos^1, P_2:pos^1, GM_1:closed^1, GM_2:open^2\}$\\
 and\\ 
$t_2 =  \{P_1:pos^2, P_2:pos^2, GM_1:closed^1, GM_2:open^2\}$,\\ erroneously omitting a test case including $pos^3$. Once the tester submits the test plan, the two tests are marked as  \emph{validated} by the tester. At this point the tester's mistake may be discovered, as pairwise coverage is not achieved: e.g., the interactions $\{P_1:pos_1,P_2:pos_2\}$ and $\{P_1:pos_3,P_2:pos_3\}$ remain uncovered. This can be either due to the fact that the tester considered non-executable tests as possible (as in the first interaction), or forgot to add some tests (as in the second interaction). 

Our framework provides a human-oriented solution to this kind of problems: 
the corresponding query from the framework could prompt the tester to either update the logical condition with $P_1 = P_2$ (thus removing the interaction $\{P_1:pos_1,P_2:pos_2\}$ from coverage requirements) or extend the test plan with the test 
 $\{P_1:pos^3, P_2:pos^3, GM_1:close^1, GM_2:open^2\}$.

\section{Conclusions and Future Work}
\label{sec:conclusion}

In this paper, we presented our ongoing work on human-centred testing of CPSs. 
We integrate the ideas of  refinement-based development  and the agile combinatorial test design, 
a human-centred methodology, which takes into account the human tester's possible mistakes and supports
revision and refinement. 
 We also aim at increasing of the readability and understandability of tests, 
 to conform with the ideas of human-oriented software development, cf. \cite{spichkova2013we,spichkova2013design,hffm_spichkova}. 
The suggested approach can significantly increase efficiency of testing of CPSs and reduce the testing cost and time by following the agile paradigm and providing an interactive support to the tester. 

While in agile CTD an iterative process concerns only the interaction between a model and a test plan, 
in the current framework we have several inter-related levels of abstraction and their corresponding test plans. 
Incorporating an human-centred iterative process of refinement into the framework leads to a number of interesting questions. 
How does the refinement of the system model on one of the meta-levels of abstraction (abstract, virtual, cyber-physical) affect the other levels? 
How does a refinement of a test plan for one of the levels affect the other test plans? 
How can ``propagation" of errors and of their correction be formalised? 
These questions provide us the main directions for our future work. 

A further future work direction is an implementation of a tool prototype for the proposed framework. To this end we plan to extend the environment of IBM Functional Coverage Unified Solution (IBM FoCuS, cf. \cite{6227245,Wojciak:2014}), which is a tool for test-oriented system modelling, which main functions are model based test planning and functional
coverage analysis.

\balance 

\begin{thebibliography}{10}

\bibitem{blech2014cyber}
J.~O. Blech, M.~Spichkova, I.~Peake, and H.~Schmidt.
\newblock Cyber-virtual systems: Simulation, validation \& visualization.
\newblock In {\em 9th International Conference on Evaluation of Novel
  Approaches to Software Engineering (ENASE 2014)}, 2014.

\bibitem{broy_refinement}
M.~Broy.
\newblock Compositional refinement of interactive systems.
\newblock {\em ACM}, 44(6):850--891, 1997.

\bibitem{broy_janus}
M.~Broy.
\newblock {Service-oriented Systems Engineering: Specification and design of
  services and layered architectures. {T}he {JANUS} {A}pproach}.
\newblock {\em Engineering Theories of Software Intensive Systems}, pages
  47--81, 2005.

\bibitem{Dhillon}
B.~Dhillon.
\newblock {\em Engineering Usability: Fundamentals, Applications, Human
  Factors, and Human Error}.
\newblock American Scientific Publishers, 2004.

\bibitem{Drath2008}
R.~Drath, P.~Weber, and N.~Mauser.
\newblock An evolutionary approach for the industrial introduction of virtual
  commissioning.
\newblock In {\em IEEE International Conference on Emerging Technologies and
  Factory Automation (ETFA)}, pages 5--8, 2008.

\bibitem{farchi2013using}
E.~Farchi, I.~Segall, and R.~Tzoref-Brill.
\newblock Using projections to debug large combinatorial models.
\newblock In {\em IEEE 6th International Conference o Software Testing,
  Verification and Validation Workshops (ICSTW)}, pages 311--320. IEEE, 2013.

\bibitem{farchi2014combinatorial}
E.~Farchi, I.~Segall, R.~Tzoref-Brill, and A.~Zlotnick.
\newblock Combinatorial testing with order requirements.
\newblock In {\em IEEE 7th International Conference on Software Testing,
  Verification and Validation Workshops (ICSTW)}, pages 118--127. IEEE, 2014.

\bibitem{feilkas2009top}
M.~Feilkas, A.~Fleischmann, F.~H{\"o}lzl, C.~Pfaller, K.~Scheidemann,
  M.~Spichkova, and D.~Trachtenherz.
\newblock {A top-down methodology for the development of automotive software}.
\newblock Technical Report TUM-I0902, TU M{\"u}nchen, 2009.

\bibitem{feilkas2011refined}
M.~Feilkas, F.~H{\"o}lzl, C.~Pfaller, S.~Rittmann, B.~Sch{\"a}tz, W.~Schwitzer,
  W.~Sitou, M.~Spichkova, and D.~Trachtenherz.
\newblock {A Refined Top-Down Methodology for the Development of Automotive
  Software Systems - The KeylessEntry System Case Study}.
\newblock Technical Report TUM-I1103, TU M{\"u}nchen, 2011.

\bibitem{hazzan2014agile}
O.~Hazzan and Y.~Dubinsky.
\newblock The agile manifesto.
\newblock In {\em Agile Anywhere}, pages 9--14. Springer International
  Publishing, 2014.

\bibitem{hellmann2012agile}
T.~D. Hellmann, A.~Sharma, J.~Ferreira, and F.~Maurer.
\newblock Agile testing: Past, present, and future--charting a systematic map
  of testing in agile software development.
\newblock In {\em Agile Conference (AGILE), 2012}, pages 55--63. IEEE, 2012.

\bibitem{Kuhn:2004}
D.~R. Kuhn, D.~R. Wallace, and A.~M. Gallo, Jr.
\newblock Software fault interactions and implications for software testing.
\newblock {\em IEEE Transactions on Software Engineering}, 30(6):418--421, June
  2004.

\bibitem{Liu2015ASWEC1}
H.~Liu, M.~Spichkova, H.~Schmidt, T.~Sellis, and M.~Duckham.
\newblock Spatio-temporal architecture-based framework for testing services in
  the cloud.
\newblock In {\em 24th Australasian Software Engineering Conference (ASWEC
  2015)}, 2015.

\bibitem{Liu2015ASWEC2}
H.~Liu, M.~Spichkova, H.~Schmidt, A.~Ulrich, H.~Sauer, and J.~Wieghardt.
\newblock Efficient testing based on logical architecture.
\newblock In {\em 24th Australasian Software Engineering Conference (ASWEC
  2015)}, 2015.

\bibitem{makris2012virtual}
S.~Makris, G.~Michalos, and G.~Chryssolouris.
\newblock Virtual commissioning of an assembly cell with cooperating robots.
\newblock {\em Advances in Decision Sciences}, 2012, 2012.

\bibitem{mioch2011selecting}
T.~Mioch, J.-P. Osterloh, and D.~Javaux.
\newblock Selecting human error types for cognitive modelling and simulation.
\newblock In {\em Human modelling in assisted transportation}, pages 129--138.
  Springer, 2011.

\bibitem{Nie:2011}
C.~Nie and H.~Leung.
\newblock A survey of combinatorial testing.
\newblock {\em ACM Comput. Surv.}, 43(2):11:1--11:29, Feb. 2011.

\bibitem{RedmillRajan}
F.~Redmill and J.~Rajan.
\newblock {\em Human factors in safety-critical systems}.
\newblock {Butterworth-Heinemann}, 1997.

\bibitem{Rumpe2006Agile}
B.~Rumpe.
\newblock Agile test-based modeling.
\newblock In {\em Proceedings of the 2006 International Conference on Software
  Engineering Research \& Practice (SERP)}. CSREA Press, 2006.

\bibitem{Sapienza2690}
G.~Sapienza, I.~Crnkovic, and T.~Seceleanu.
\newblock Towards a methodology for hardware and software design separation in
  embedded systems.
\newblock In {\em Proc. of the ICSEA}, pages 557--562. IARIA, 2012.

\bibitem{6227245}
I.~Segall and R.~Tzoref-Brill.
\newblock Interactive refinement of combinatorial test plans.
\newblock In {\em Software Engineering (ICSE), 2012 34th International
  Conference on}, pages 1371--1374, 2012.

\bibitem{segall2012common}
I.~Segall, R.~Tzoref-Brill, and A.~Zlotnick.
\newblock Common patterns in combinatorial models.
\newblock In {\em Proceedings of the IEEE Fifth International Conference on
  Software Testing, Verification and Validation (ICST)}, pages 624--629. IEEE,
  2012.

\bibitem{spichkova2008refinement}
M.~Spichkova.
\newblock Refinement-based verification of interactive real-time systems.
\newblock {\em Electronic Notes in Theoretical Computer Science}, 214:131--157,
  2008.

\bibitem{spichkova2010architecture}
M.~Spichkova.
\newblock Architecture: Methodology of decomposition.
\newblock Technical Report TUM-I1018, TU M{\"u}nchen, 2010.

\bibitem{ArchReqDecRef}
M.~Spichkova.
\newblock {Architecture: Requirements + Decomposition + Refinement}.
\newblock {\em Softwaretechnik-Trends}, 31:4, 2011.

\bibitem{hffm_spichkova}
M.~Spichkova.
\newblock {Human Factors of Formal Methods}.
\newblock In {\em {IADIS Interfaces and Human Computer Interaction 2012}}. IHCI
  2012, 2012.

\bibitem{spichkova2013design}
M.~Spichkova.
\newblock Design of formal languages and interfaces: "formal" does not mean
  "unreadable".
\newblock In {\em Emerging Research and Trends in Interactivity and the
  Human-Computer Interface}. IGI Global, 2013.

\bibitem{Spichkova_Campetelli2012}
M.~Spichkova and A.~Campetelli.
\newblock Towards system development methodologies: From software to
  cyber-physical domain.
\newblock In {\em First International Workshop on Formal Techniques for
  Safety-Critical Systems}, 2012.

\bibitem{VerisoftXT_FMDS}
M.~Spichkova, F.~Hošlzl, and D.~Trachtenherz.
\newblock Verified system development with the autofocus tool chain.
\newblock In {\em 2nd Workshop on Formal Methods in the Development of
  Software}, WS-FMDS, 2012.

\bibitem{spichkova2015human}
M.~Spichkova, H.~Liu, M.~Laali, and H.~W. Schmidt.
\newblock Human factors in software reliability engineering.
\newblock {\em Workshop on Applications of Human Error Research to Improve
  Software Engineering (WAHESE2015)}, 2015.

\bibitem{spichkova2015SAGRA}
M.~Spichkova, H.~Liu, and H.~Schmidt.
\newblock Towards quality-oriented architecture: Integration in a global
  context.
\newblock In {\em Proceedings of the 2015 European Conference on Software
  Architecture Workshops}, page~64. ACM, 2015.

\bibitem{issec2013spichkova}
M.~Spichkova, H.~Schmidt, and I.~Peake.
\newblock {From abstract modelling to remote cyber-physical
  integration/interoperability testing}.
\newblock In {\em {Improving Systems and Software Engineering Conference}},
  2013.

\bibitem{spichkova2013we}
M.~Spichkova, X.~Zhu, and D.~Mou.
\newblock Do we really need to write documentation for a system?
\newblock In {\em International Conference on Model-Driven Engineering and
  Software Development (MODELSWARD'13)}, 2013.

\bibitem{Tai}
K.-C. Tai and Y.~Lei.
\newblock A test generation strategy for pairwise testing.
\newblock {\em IEEE Transactions on Software Engineering}, 28(1):109--111,
  2002.

\bibitem{talby2006agile}
D.~Talby, A.~Keren, O.~Hazzan, and Y.~Dubinsky.
\newblock Agile software testing in a large-scale project.
\newblock {\em IEEE Software}, 23(4):30--37, 2006.

\bibitem{Turk2005Agile}
D.~Turk, R.~B. France, and B.~Rumpe.
\newblock Assumptions underlying agile software development processes.
\newblock {\em Journal of Database Management}, 16:62--87, 2005.

\bibitem{Walia2013}
G.~Walia and J.~Carver.
\newblock Using error information to improve software quality.
\newblock In {\em IEEE International Symposium on Software Reliability
  Engineering Workshops (ISSREW)}, pages 107--107, 2013.

\bibitem{Wojciak:2014}
P.~Wojciak and R.~Tzoref-Brill.
\newblock System level combinatorial testing in practice -- the concurrent
  maintenance case study.
\newblock In {\em Proceedings of the 2014 IEEE International Conference on
  Software Testing, Verification, and Validation}, ICST '14, pages 103--112.
  IEEE Computer Society, 2014.

\bibitem{Zamansky2015HOFM}
A.~Zamansky and E.~Farchi.
\newblock Helping the tester get it right: Towards supporting agile
  combinatorial test design.
\newblock In {\em 2nd Human-Oriented Formal Methods workshop (HOFM 2015)},
  2015.

\bibitem{zhang2014introduction}
J.~Zhang, Z.~Zhang, and F.~Ma.
\newblock Introduction to combinatorial testing.
\newblock In {\em Automatic Generation of Combinatorial Test Data}, pages
  1--16. Springer, 2014.

\end{thebibliography}

\end{document}